\documentclass[12pt]{article}
\usepackage{epsf}
\def\dst{\displaystyle}
\def\f{\frac}

\def\m{\mu}

\def\e{\varepsilon}
\def\eps{\varepsilon}
\def\be{\begin{equation}}
\def\ee{\end{equation}}
\def\bea{\begin{eqnarray}}
\def\eea{\end{eqnarray}}

\def\gsim{\raisebox{-0.6ex}{$\stackrel{\textstyle >}{\sim}$}}

\begin{document}
\begin{titlepage}
\begin{center}
{\Large \bf William I. Fine Theoretical Physics Institute \\
University of Minnesota \\}
\end{center}
\vspace{0.2in}
\begin{flushright}
FTPI-MINN-08/31 \\
UMN-TH-2713/08 \\
August 2008 \\
\end{flushright}
\vspace{0.3in}
\begin{center}
{\Large \bf The spontaneous breaking of a metastable string
\\}
\vspace{0.2in}
{\bf A. Monin \\}
School of Physics and Astronomy, University of Minnesota, \\ Minneapolis, MN
55455, USA, \\
and \\
{\bf M.B. Voloshin  \\ }
William I. Fine Theoretical Physics Institute, University of
Minnesota,\\ Minneapolis, MN 55455, USA \\
and \\
Institute of Theoretical and Experimental Physics, Moscow, 117218, Russia
\\[0.2in]
\end{center}

\begin{abstract}

We consider the phase transition of a string with tension $\eps_1$ to a string
with a smaller tension $\eps_2$. The transition proceeds through quantum
tunneling, and we calculate in arbitrary number of dimensions the
pre-exponential factor multiplying the leading semiclassical exponential
expression for the rate of the process. At $\eps_2=0$ the found formula for the
decay rate also describes a break up of a metastable string into two pieces.

\end{abstract}

\end{titlepage}

\section{Introduction}
String-like configurations arise in a great number of actual systems, such as
superconductors and polymers, in theoretical models of non-Abelian dynamics,
such
as models of QCD confinement, and as topological defects in models with
spontaneous breaking of gauge or global symmetries. In some of these situations
the string configurations are not absolutely stable, but are rather metastable
with respect to either a complete breaking, or a break up with a string of lower
tension
emerging instead of the initial string. The former situation is relevant e.g.
for a break up of a QCD string with formation of a quark - antiquark pair, or a
formation of a monopole - antimonopole pair\cite{Vilenkin}, and also in a whole
class of theories with spontaneous symmetry breaking\cite{Preskill}. The latter
situation involving a phase transition between states of a string with different
tension is found e.g. in Abelian Higgs models embedded in non-Abelian
theories\cite{Shifman}. It is clear that the latter case is more general, since
the breaking of a string into `nothing' can be considered as a transition into a
string with zero tension.

The transition of a metastable string with the tension $\eps_1$ to a string with
a lower tension $\eps_2$ is quite analogous to decay of metastable
vacuum\cite{vko,Coleman}, or the Schwinger process of creation of pairs of
charged particles by electric field\cite{Schwinger}. Indeed, if a piece of
length $\ell$ is converted into the lower phase, the gain in the energy is
$(\eps_1 - \eps_2) \, \ell$. The barrier that inhibits the process is created by
the energy $\mu$ associated with the interface between the phases of the string,
e.g. the mass of the monopole in the examples of Ref.\cite{Vilenkin} or
\cite{Shifman}. The transition involves two such interfaces, so that the
`length' energy gain exceeds the barrier energy $2 \mu$ only starting from a
critical size of the `true' phase $\ell_c=2 \mu/(\eps_1 - \eps_2)$. Once a
critical piece has nucleated due to tunneling, the additional energy gain is
spent on acceleration of the ends of this piece in its expansion, which
eventually converts the whole length of the initial string.  Thus the
probability of the transition is given by the rate of nucleation of the critical
gaps in the initial string. This rate per unit time and per unit length of the
metastable string is given by the semiclassical
expression\cite{Vilenkin,Preskill,Shifman}
\be
{d\Gamma \over d \ell} =C \, \exp \left ( - \, {\pi \, \mu^2 \over
\eps_1-\eps_2} \right )~,
\label{g0}
\ee
where $C$, the pre-exponential factor, is the main subject of the calculation in
this paper. The semiclassical expression is formally applicable as long as the
exponential power, determined by the action on the tunneling trajectory is
large, which requires
\be
\mu^2 \gg \eps_1-\eps_2~.
\label{muineq}
\ee

The pre-exponential factor $C$ is found from a calculation of the path integral
over small deviations from the semiclassical tunneling trajectory\cite{Callan}.
The result of such calculation in a (1+1) dimensional theory, can be readily
copied from the corresponding expressions in equivalent (1+1) dimensional
problems: either that of pair creation in electric field\cite{Schwinger}, or of
false vacuum decay\cite{Stone,ks,mv85},
\be
C_{d=2}= {\eps_1-\eps_2 \over 2 \pi}~.
\label{c2}
\ee
However, the equivalence of the string transition to already solved problems
does not hold for string in more than two dimensions. The loss of the relation
to the false vacuum decay is due to the obvious difference in the dimensionality
of the relevant objects, while the loss of equivalence to the Schwinger process
is due to a different dependence of the action on the deviations of the
tunneling trajectory in the transverse directions. Indeed small deviations of
the particles in the Schwinger process perpendicular to the external electric
field involve only the particles themselves, while a transverse displacement of
an end of a string involves both the `particle' (of the mass $\mu$) and an
adjacent part of the string. In this paper we calculate the relevant path
integral and find the pre-exponential coefficient for arbitrary number of
space-time dimensions $d$. Our final result for the transition rate reads as
\be
{d \Gamma \over d \ell}={\eps_1-\eps_2 \over 2 \pi} \, \left [ F \left ( {\eps_2
\over \eps_1} \right ) \right ]^{d-2} \, \exp \left ( -
\, {\pi \, \mu_R^2 \over \eps_1-\eps_2} \right )~,
\label{gf}
\ee
with $\mu_R$ being the renormalized value of the mass that includes the effects
of the
motion of the adjacent part of the string, and $F$ is the factor contributed in
the rate by each of the $(d-2)$ transverse dimensions:
\be
F \left ( {\eps_2 \over \eps_1} \right ) = \sqrt{\eps_1+\eps_2 \over 2 \,
\eps_1} \,\, {\rm \Gamma} \! \left
({\eps_1+\eps_2 \over  \eps_1 - \eps_2} +1 \right ) \, \left ( {\eps_1-\eps_2
\over  \eps_1 + \eps_2} \right )^{\eps_1+\eps_2 \over  \eps_1 - \eps_2} \, \exp
\left ( {\eps_1+\eps_2 \over  \eps_1 - \eps_2} \right )  \left ( 2 \pi \,
{\eps_1+\eps_2 \over  \eps_1 - \eps_2} \right )^{-1/2}~,
\label{ff}
\ee
where ${\rm \Gamma}(z)$ is the standard Gamma function. The expression  given by
Eqs.(\ref{gf}) and (\ref{ff}) thus replaces the previous estimates\cite{Shifman}
of the pre-exponential factor in the transition rate.

In what follows we present our calculation starting with briefly going over the
Euclidean-space path integral formulation of the tunneling problem and
introducing the string action and the relevant variables in Sec.~2. We then
consider the separation of the variables in Sec.~3 and introduce a
regularization procedure in Sec.~4. The actual calculation of the path integrals
is presented in Sec.~5 and in Sec.~6 we consider the renormalization of the mass
$\mu$, associated with the interface between the strings, by the motion of
adjacent pieces of the string. Finally, in Sec.~7 we assemble our result for the
transition rate and discuss the applicability and the behavior of the found
formula as well as some specific points of the calculation.

\section{Euclidean-space formulation and the relevant variables}
The tunneling trajectory can be described in the Euclidean space by a
configuration called the `bounce'\cite{Coleman}, which is a solution to
(Euclidean) classical equations of motion.
The general expression for the effective Euclidean space action for the string
with the two considered phases can be written in the familiar Nambu-Goto form:
\be
S=\mu \, P + \eps_1 \, A_1 + \eps_2 \, A_2~,
\label{a0}
\ee
where $A_1$ and $A_2$ are the areas of the world sheet for the two phases, and
$P$ (the perimeter) is the length of the world line for the interface between
them.

\begin{figure}[ht]
  \begin{center}
    \leavevmode
    \epsfxsize=12cm
    \epsfbox{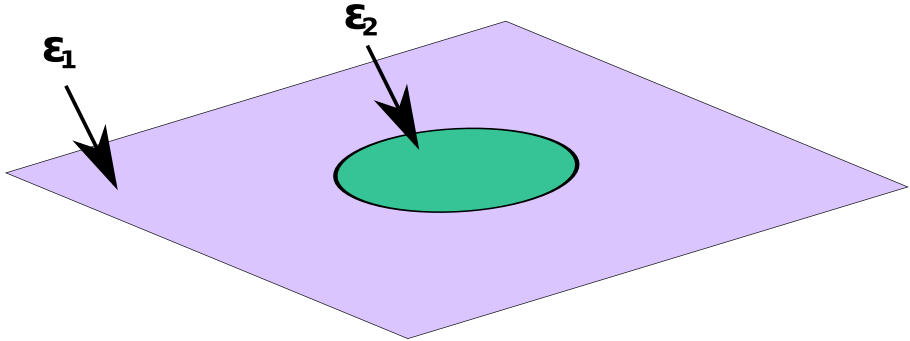}
    \caption{The bounce configuration, describing the semiclassical tunneling
trajectory.}
  \end{center}
\end{figure}

The action (\ref{a0})  is an effective low-energy expression in the sense that
it only describes the `stringy' variables and is applicable as long as the
effects of thickness of the string and of its internal structure can be
neglected.  Denoting $M_0$ the mass scale at which such approach becomes invalid
(e.g. the thickness of the string $r_0 \sim 1/M_0$), one can write the condition
for the applicability of the effective action (\ref{a0}) in terms of the length
scale, $\ell \gg 1/M_0$, and the momentum scale $k \ll M_0$.
Assuming that the initial very long string with tension $\eps_1$ is located
along the $x$ axis, one can readily find that the action (\ref{a0}) has a
nontrivial stationary configuration, the bounce, namely, that of a disk in the
(t,x) plane occupied by the phase $2$, as shown in Fig.1, with the radius
\be
R={\mu \over  \eps_1-\eps_2}~,
\label{rc}
\ee
which is the radius (one half of the length) of the critical gap. The difference
between the action (\ref{a0}) on this configuration and on the trivial one is
exactly the expression for the exponential power in Eq.(\ref{g0}), and the
condition for applicability of the effective action (\ref{a0}) requires
\be
M_0 \, R = {M_0 \, \mu \over  \eps_1-\eps_2} \gg 1~.
\label{rineq}
\ee
Generally one also has $\mu \gsim M_0$, and for the strings in weakly coupled
theories $\mu \gg M_0$, so that the power in the exponent in Eq.(\ref{g0}) is
large, which justifies a semiclassical treatment.

The probability of the transition is determined\cite{Stone,Coleman,Callan} by
(the imaginary part of) the ratio of the path integrals ${\cal Z}_{12}$ and
${\cal Z}_1$ calculated with the action (\ref{a0}) around respectively the
bounce configuration and around the initial flat string:
\be
{d \Gamma \over d \ell}= {1 \over XT} \, \mathrm{Im}\f{{\cal
Z}_{12}}{{\cal Z}_{1  }}~.
\label{rz}
\ee
It can also be reminded that, as explained in great detail in Ref.\cite{Callan},
that the imaginary part of ${\cal Z}_{12}$ arises from one negative mode at the
bounce configuration, and that due to two translational zero modes the
numerator in Eq.(\ref{rz}) is proportional to the total space time area $XT$
in the $(t,x)$ plane occupied by the string, so that the finite quantity is the
transition probability per unit time (the rate) and per unit length of the
string.

In order to evaluate the relevant path integrals with the pre-exponential
accuracy we use the cylindrical coordinates, with $r$ and $\theta$ being the
polar variables in the $(t,x)$ plane (of the bounce), and $z$ being the
transverse coordinate. We consider only one transverse coordinate, since the
effect of each of the extra dimensions factorizes, so that the corresponding
generalization is straightforward. We further assume, for definiteness, that the
space-time boundary in the $(t,x)$ plane is a circle of large radius $L$, where
the
boundary condition for the string is $z(r=L)=0$. The small deviations of the
string configuration from the bounce, illustrated in Fig.2, can be parametrized
by the radial ($f$) and transverse ($\zeta$) shifts of the boundary between the
string phases:
\be
r(\theta) = R + f(\theta)\,~~~~z(\theta)=\zeta(\theta)~,
\label{bv}
\ee
and the variations of the surfaces of the two string phases: $z_1(r,\theta)$ and
$z_2(r,\theta)$, where, naturally,
\be
z_1(R,\theta)= z_2(R,\theta)=\zeta(\theta)~.
\label{bc}
\ee

\begin{figure}[ht]
  \begin{center}
    \leavevmode
    \epsfxsize=12cm
    \epsfbox{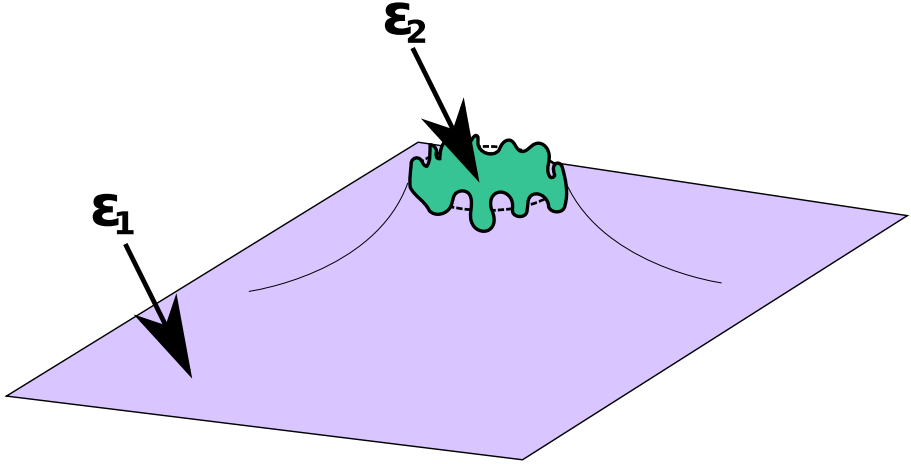}
    \caption{Fluctuations around the bounce configuration.}
  \end{center}
\end{figure}

In terms of these variables the action (\ref{a0}) can be written in the
quadratic approximation in the deviations from the bounce as
\bea
&&S_{12}=\eps_1 \, \pi \, L^2 + {\pi \, \mu^2 \over \eps_1- \eps_2} +
{\eps_1-\eps_2 \over 2} \, \int d\theta \, \left (\dot{\zeta}^2+\dot{f}^2-f^2
\right )+
\nonumber \\
&& {\eps_1 \over 2} \, \int_R^L r drd\theta \,\left ( z_1'\,^2+ {\dot{z_1}^2
\over r^2} \right ) + {\eps_2 \over 2} \, \int_0^R r drd\theta \,\left (
z_2'\,^2+ {\dot{z_2}^2 \over r^2} \right )~,
\label{s12}
\eea
where the primed and dotted symbols stand for the derivatives with respect to
$r$ and $\theta$ correspondingly.

Finally, the action around a flat initial string configuration in the quadratic
approximation takes the form
\be
S_1 = \eps_1 \, \pi L^2 +{\eps_1 \over 2} \, \int_0^L r drd\theta \,\left (
z'\,^2+ {\dot{z}^2 \over r^2} \right )~,
\label{s1}
\ee
with $z(r,\theta)$ parametrizing small deviations of the string in the
transverse direction.

\section{Separating variables in the path integrals}
One can readily see that in the quadratic part of the action (\ref{s12}) the
`longitudinal' variation of the bounce boundary in the $(t,x)$ plane, described
by the function $f(\theta)$ completely decouples from the rest of the variables.
This implies that the path integral over $f$ can be considered independently of
the integration over other variables and that it enters as a factor in ${\cal
Z}_{12}$. On the other hand it is this integral that provides the imaginary part
to the partition function, and it is also proportional to the total space-time
area $XT$. Moreover, this path integral is identical to the one entering the
problem of false vacuum decay in (1+1) dimensions and we can directly apply the
result of that calculation\cite{mv85}:
\be
{1 \over XT} \, \mathrm{Im} \int {\cal D} f \, \exp \left [ - {\eps_1-\eps_2
\over 2} \, \int_0^{2\pi} \, \left ( \dot{f}^2 - f^2 \right ) \, d\theta \right]
=  {\eps_1-\eps_2 \over 2 \pi}~.
\label{dim2r}
\ee
The expression for the transition rate thus can be written in the form
\be
{d \Gamma \over d \ell} = {\eps_1-\eps_2 \over 2 \pi} \, \exp \left ( -{\pi
\,\mu^2 \over \eps_1-\eps_2} \right ) \, \, {{\tilde {\cal Z}}_{12} \over {\cal
Z}_1} ~,
\label{g2}
\ee
with the path integral ${\tilde {\cal Z}}_{12}$ running only over the transverse
variables $\zeta$, $z_1$ and $z_2$,
\be
{\tilde {\cal Z}}_{12} = \int {\cal D}\zeta \, {\cal D}z_1 \, {\cal D} z_2 \,
\exp \left (- {\tilde S}_{12} \right)
\label{tz12}
\ee
and involving only the quadratic in these variables part of the action
(\ref{s12}),
\be
{\tilde S}_{12} = {\eps_1-\eps_2 \over 2} \, \int d\theta \, \dot{\zeta}^2 +
{\eps_1 \over 2} \, \int_R^L r drd\theta \,\left ( z_1'\,^2+ {\dot{z_1}^2 \over
r^2} \right ) + {\eps_2 \over 2} \, \int_0^R r drd\theta \,\left ( z_2'\,^2+
{\dot{z_2}^2 \over r^2} \right )~.
\label{ts12}
\ee
In the same quadratic approximation the flat string partition function ${\cal
Z}_1$ is given by
\be
{\cal Z}_1= \int {\cal D} z \, \exp \left (- S_1 \right )
\label{z1}
\ee
with $S_1$ given by Eq.(\ref{s1}) and the integral running over all the
functions vanishing at the space-time boundary: $z(L, \theta)=0$.

At this point there still is a coupling in the path integral (\ref{tz12})
between the bulk variables $z_1$, $z_2$ and the boundary variable $\zeta$
arising from the boundary conditions (\ref{bc}). This however is a simple issue
which is resolved by a straightforward shift of the integration variables $z_1$
and $z_2$. Namely, we write
\be
z_1(r, \theta)=z_{1c}(r, \theta) + z_{1q}(r, \theta)~,~~~~z_2(r,
\theta)=z_{2c}(r, \theta) + z_{2q}(r, \theta)~,
\label{zcq}
\ee
where $z_{1q}$ and $z_{2q}$ are the new integration variables in ${\tilde {\cal
Z}}_{12}$ and these functions satisfy zero boundary conditions,
\be
z_{1q}(R, \theta)=z_{2q}(R, \theta)=z_{1q}(L, \theta)=z_{2q}(L, \theta)=0~,
\label{zqbc}
\ee
while $z_{1c}$ and $z_{2c}$ are fixed (for a fixed $\zeta(\theta)$) functions
satisfying the boundary conditions
\be
z_{1c}(R, \theta)=z_{2c}(R, \theta)=\zeta(\theta)~, z_{1c}(L, \theta)=0~,
\label{zcbc}
\ee
which are also harmonic, i.e. satisfying the Laplace equation $\Delta z=0$.

After the specified shift of the variables we arrive at the expression for the
action (\ref{ts12}) in which the bulk and the boundary degrees of freedom are
fully separated:
\bea
{\tilde S}_{12} &=& {\eps_1-\eps_2 \over 2} \, \int d\theta \, \dot{\zeta}^2 + R
\, \int d \theta  \,  \left ( {\eps_2 \over 2} {\partial_r} z_{2c} \left .
\right |_{r=R} - {\eps_1 \over 2} {\partial_r} z_{1c} \left . \right |_{r=R}
\right ) \, \zeta
\nonumber \\
&-& {\eps_1 \over 2} \, \int d^2r \, z_{1q} \Delta z_{1q} - {\eps_2 \over 2} \,
\int d^2r \, z_{2q} \Delta z_{2q}~.
\label{tss}
\eea
Clearly, the boundary terms in the first line here, arising from the integration
by parts, depend only on the transverse shift of the boundary $\zeta(\theta)$.
The partition function ${\tilde {\cal Z}}_{12}$ can thus be written as a product
of the `boundary' and the `bulk' terms:
\be
{\tilde {\cal Z}}_{12} = {\cal Z}_{12 (\rm boundary)} \, {\cal Z}_{12 (\rm
bulk)}~,
\label{z12bb}
\ee
with the ${\cal Z}_{12 (\rm bulk)}$ being given by the path integral over the
bulk variables $z_{1q}$ and $z_{2q}$ only :
\be
{\cal Z}_{12 (\rm bulk)}=\int {\cal D} z_{1q} \,  {\cal D} z_{q} \, \exp \left (
{\eps_1 \over 2} \, \int d^2r \, z_{1q} \Delta z_{1q} + {\eps_2 \over 2} \, \int
d^2r \, z_{2q} \Delta z_{2q} \right )~,
\label{z12bk}
\ee
and the boundary term
\be
{\cal Z}_{12 (\rm boundary)}=\int {\cal D}\zeta \, \exp \left [ - {\eps_1-\eps_2
\over 2} \, \int d\theta \, \dot{\zeta}^2 - R \, \int d \theta  \,  \left (
{\eps_2 \over 2} {\partial_r} z_{2c} \left . \right |_{r=R} - {\eps_1 \over 2}
{\partial_r} z_{1c} \left . \right |_{r=R} \right ) \, \zeta \right]
\label{z12by}
\ee
involving integration over only the boundary values.

The subsequent calculation of the ratio of the partition functions in
Eq.(\ref{g2}) can in fact be reduced to a calculation of ${\cal Z}_{12 (\rm
boundary)}$ only. In order to achieve this reduction one should organize the
partition function ${\cal Z}_1$ in the denominator of Eq.(\ref{g2}) in a form
similar to Eq.(\ref{z12bb}) as follows. Although the flat string configuration
makes no reference to a circle of the radius $R$, the partition function
${\cal Z}_1$ can be calculated by first fixing the transverse variable $z$ at
$r=R$: $z(R, \theta)= \zeta(\theta)$ and separating the integration over the
bulk variables. Then the flat string partition function factorizes in the form
similar to Eq.(\ref{z12bb}):
\be
{\tilde {\cal Z}}_{1} = {\cal Z}_{1 (\rm boundary)} \, {\cal Z}_{1 (\rm bulk)}~,
\label{z1bb}
\ee
with ${\cal Z}_{1 (\rm bulk)}$ given a similar path integral as in
Eq.(\ref{z12bk}),
\be
{\cal Z}_{1 (\rm bulk)}=\int {\cal D} z_{1q} \,  {\cal D} z_{q} \, \exp \left (
{\eps_1 \over 2} \, \int d^2r \, z_{1q} \Delta z_{1q} + {\eps_1 \over 2} \, \int
d^2r \, z_{2q} \Delta z_{2q} \right )~,
\label{z1bk}
\ee
where, as in Eq.(\ref{z12bk}),  $z_{1q}$ and $z_{2q}$ are respectively the outer
(i.e. at $r > R$) and the inner ($r < R$) transverse fluctuations with zero
boundary conditions. The difference in the coefficient in the expressions
(\ref{z12bk}) and (\ref{z1bk}) for the contribution of the inner part, $\eps_2$
vs. $\eps_1$, is not essential, since the overall coefficient of the quadratic
part of the action is absorbed in the measure of integration, as can be seen by
rescaling to the corresponding canonically normalized variables $\phi =
\sqrt{\eps} \, z$.

One therefore finds that ${\cal Z}_{1 (\rm bulk)} = {\cal Z}_{12 (\rm bulk)}$,
and the ratio of the partition functions in Eq.(\ref{g2}) is in fact determined
by the ratio of the boundary terms.

\section{Regularization}
The boundary factor ${\cal Z}_{1 (\rm boundary)}$ for the flat string is
somewhat different from the one given by Eq.(\ref{z12by}) and reads as
\be
{\cal Z}_{1 (\rm boundary)}=\int {\cal D}\zeta \, \exp \left [ -R \,{\eps_1
\over 2} \, \int d \theta \,  \left ({\partial_r} z_{2c} \left . \right |_{r=R}
-  {\partial_r} z_{1c} \left . \right |_{r=R} \right ) \, \zeta \right]~,
\label{z1by}
\ee
where the functions $z_{1c}$ and $z_{2c}$ are defined in the same way as in
Eq.(\ref{z12by}).

The latter functions can be readily found by expanding the boundary function
$\zeta(\theta)$ in angular harmonics:
\be
\zeta(\theta)={a_0 \over \sqrt{2\pi}} + {1 \over \sqrt{\pi}}\sum_{n=1}^\infty
\left [ a_n \, \cos( n \theta)+b_n \, \sin(n \theta) \right ]
\label{ft}
\ee
with $a_n$ and $b_n$ being the amplitudes of the harmonics. Then at $n \neq 0$
the harmonics of the discussed functions are found as
\be
z_{1c}^{(n)}(r, \theta)= {1 \over \sqrt{\pi}} \,\left [ a_n \, \cos( n
\theta)+b_n \, \sin(n \theta) \right ] \, {R^n \over r^n}\,, ~~z_{2c}^{(n)}(r,
\theta)= {1 \over \sqrt{\pi}} \, \left [ a_n \, \cos( n \theta)+b_n \, \sin(n
\theta) \right ] \, {r^n \over R^n}~,
\label{hn}
\ee
and for n=0 these are given by
\be
z_{1c}^{(0)}(r, \theta) = a_0 \, {\ln (r/L) \over \ln (R/L)}~, ~~~~
z_{2c}^{(0)}(r, \theta) =a_0~.
\label{h0}
\ee
Substituting these expressions for the harmonics in the equations (\ref{z12by})
and (\ref{z1by}) and performing the Gaussian integration over the amplitudes
$a_n$ and $b_n$ we find the  boundary factors in the form
\be
{\cal Z}_{12 (\rm boundary)}={\cal N} \,  \sqrt{\eps_1 \, \ln {L \over R}} \,
\prod_{n=1}^\infty {1 \over (\eps_1-\eps_2) \, n^2 + (\eps_1 + \eps_2) \, n}~
\label{z12by1}
\ee
and
\be
{\cal Z}_{1 (\rm boundary)}={\cal N} \,  \sqrt{\eps_1 \, \ln {L \over R}} \,
\prod_{n=1}^\infty {1 \over 2 \, \eps_1 \, n}
\label{z1by1}
\ee
with ${\cal N}$ being a normalization factor.

Clearly, each of the formal expressions (\ref{z12by1}) and (\ref{z1by1})
contains a divergent product, and their ratio is also ill defined, so that our
calculation requires a regularization procedure that would cut off the
contribution of harmonics with large $n$. A regularization of high harmonics is
also required on general grounds. Indeed, as previously mentioned, our
consideration using the effective string action (\ref{a0}) is only valid for
smooth deformations of the string, i.e. as long as the relevant momenta are
smaller than the mass scale $M_0$ for excitation of the internal degrees of
freedom within the thickness of the string. For an $n$-th harmonic the relevant
momentum is $k \sim n/R$ so that the applicability of the effective low energy
action requires a cutoff at $n \ll M_0R$.   In order to perform such
regularization we use the standard Pauli-Villars method and introduce a
regulator field $Z$ with negative norm and the action corresponding to the
quadratic part of the Nambu-Goto expression (\ref{a0}) for small $z$:
\be
S_R=\f{\eps_1 - \eps_2}{2} \int d\theta \dot{\zeta_R}^2+  {\eps_1 \over 2} \,
\int_{A_1} d^2r \left [ (\partial_\mu Z)^2 + M^2 \, Z^2
\right ] + {\eps_2 \over 2} \, \int_{A_2} d^2r \left [ (\partial_\mu Z)^2 + M^2
\, Z^2 \right ]
\label{sr}
\ee
with $M$ being the regulator mass, which physically should be understood as
satisfying the condition $M \ll M_0$ and still being much larger than the
relevant scale in the discussed problem, in particular $MR \gg 1$.

The regularized expression for the ratio of the boundary terms in ${\cal
Z}_{12}$ and ${\cal Z}_{1}$ thus takes the form
\be
{{\cal Z}_{12 (\rm boundary)} \over {\cal Z}_{1 (\rm boundary)}} \longrightarrow
{\cal R}=
\left [ {{\cal Z}_{12 (\rm boundary)} \over {\cal Z}^{(R)}_{12 (\rm boundary)}}
\right ] \, \left [ {{\cal Z}_{1 (\rm boundary)} \over {\cal Z}^{(R)}_{1 (\rm
boundary)}} \right ]^{-1}~,
\label{reg}
\ee
where we introduced the notation ${\cal R}$ for the regularized ratio, and the
regulator partition functions ${\cal Z}^{(R)}_{12 (\rm boundary)}$
and ${\cal Z}^{(R)}_{1 (\rm boundary)}$ are determined by the same expressions
as in Eqs.(\ref{z12by}) and (\ref{z1by}) with the `outer' and `inner' functions
$z_{1c}$ and $z_{2c}$ being replaced by their regulator counterparts $Z_{1c}$
and $Z_{2c}$ which still satisfy the boundary conditions similar to
(\ref{zcbc}):
\be
Z_{1c} (R, \theta)= Z_{2c}(R, \theta)=\zeta_R(\theta)~,
\label{zrbc}
\ee
but are the solutions of the Helmholtz rather than Laplace equation $(\Delta -
M^2) Z=0$.

The solutions of the latter equation fall off exponentially at the scale
determined by $M$, and for our purposes only the behavior near the circle $r=R$
is needed. For this reason we write the equation for the radial part of the
$n$-th angular harmonic $Z_n(r)$,
\be
Z_n''+{1 \over r} \, Z_n' -{n^2 \over r^2} \, Z_n - M^2 \, Z_n = 0~,
\label{zneq}
\ee
and parametrize the radial coordinate as $r=R+x$, and treat the parameter
$(x/R)$ as small, since the scale for the variation of the solution is $x \sim
1/\sqrt{M^2+n^2/R^2}$. This approach yields an expansion of the regulator action
associated with the boundary at $r=R$ in powers of $1/\sqrt{(MR)^2 + n^2}$. With
the accuracy required in the present calculation, the (normalized to one at
$r=R$) solution to Eq.(\ref{zneq}) is found in the first order of expansion in
$(x/R)$ as
\be
Z_n(R+x)=\left ( 1- {1 \over 2} \, {(M R)^2 \over  (MR)^2+n^2} \, {x \over R}
\right ) \, \exp \left ( -\sqrt{(MR)^2+n^2}
 \, {|x| \over R} \right )~.
\label{znsol}
\ee

Using the form of the solutions for the harmonics of the regulator field given
by Eq.(\ref{znsol}) and the expressions (\ref{z12by1}) and (\ref{z1by1}), one
can
write the regularized ratio of the boundary partition functions (\ref{reg}) as
\bea
{\cal R} &=& \sqrt{\eps_1+\eps_2 \over 2 \eps_1} \, \left [ \prod_{n=1}^\infty
{n^2 +b \, \sqrt{(MR)^2+n^2} \over n^2 + b \, n} \right ] \, \left [
\prod_{n=1}^\infty {n \over \sqrt{(MR)^2+n^2}} \right ]\, \times \nonumber \\
&&\prod_{n=1}^\infty \left \{ 1 +  {1 \over 2} \, {(MR)^2 \over \left
[(MR)^2+n^2 \right ] \, \left [ n^2 +b \, \sqrt{(MR)^2+n^2}\, \right ] } \right
\}~,
\label{crprod}
\eea
where we introduced the notation
\be
b={\eps_1+\eps_2 \over \eps_1-\eps_2}~,
\label{bdef}
\ee
and the last product factor in Eq.(\ref{crprod}) arises from the first term of
expansion in $(x/R)$ in the pre-exponential factor in Eq.(\ref{znsol})

\section{Calculating the products}
Each of the products in Eq.(\ref{crprod}) is finite at a finite $M$ and can be
calculated separately. We start with the most straightforward one, which is
directly given by an Euler's formula:
\be
\prod_{n=1}^\infty {n \over \sqrt{(MR)^2+n^2}} = \sqrt{\pi M R \over \sinh (\pi
MR)} \longrightarrow \sqrt{2 \pi M R } \, \exp \left ( - {\pi M R \over 2}
\right )~,
\label{mprod}
\ee
where the last transition corresponds to the limit $M R \gg 1$.

The other two factors in Eq.(\ref{crprod}), the first and the last, generally
depend on the relation between the parameters $b$ and $MR$, or equivalently
between $(\eps_1+\eps_2)$ and $\mu M$. We find however that the latter product
is equal to one at $MR \gg 1$ independently of $b$. In particular in the
nontrivial case of $b \ll MR$ we find
\bea
&&\ln  \prod_{n=1}^\infty \left . \left \{ 1 +  {1 \over 2} \, {(MR)^2 \over
\left [(MR)^2+n^2 \right ] \, \left [ n^2 +b \, \sqrt{(MR)^2+n^2}\, \right ] }
\right \} \right|_{MR \to \infty} \to
\nonumber \\
&&\left . \left \{ {(MR)^2 \over 2} \, \int_{n_0}^\infty dn \, \left [(MR)^2+n^2
\right ]^{-1} \,
 \left [ n^2 +b \, \sqrt{(MR)^2+n^2}\, \right ]^{-1} \right \}  \right|_{MR \to
\infty} \to 0~,
\label{f1}
\eea
where the lower limit in the integral is any number $n_0$ that is finite in the
limit $MR \to \infty$.

The dependence of the first product factor in Eq.(\ref{crprod}) on the ratio
$(\eps_1+\eps_2)/(\mu M)= b/(MR)$ is essential and we consider two limiting
cases when this ratio is much bigger than one and when it is very small. In the
former case, i.e. for $b \gg MR$, the first product in Eq.(\ref{crprod}) becomes
reciprocal of the second, and one finds
\be
{\cal R}|_{b \gg MR \gg 1}=1~.
\label{rl}
\ee
(Clearly one can also safely make the replacement $(\eps_1+\eps_2)/(2 \eps_1)
\to 1$ at $b \gg 1$.)

The behavior of ${\cal R}$ in the opposite limit, i.e. at $b \ll MR$, turns out
to be significantly more interesting. Using the Euler-Maclaurin summation
formula for the logarithm of the first product in Eq.(\ref{crprod}) we find in
the limit $MR \gg 1$ and $MR \gg b$:
\be
\prod_{n=1}^\infty {n^2 +b \, \sqrt{(MR)^2+n^2} \over n^2 + b \, n} =
{\Gamma(b+1) \over 2 \pi \, \sqrt{b MR}} \, \exp \left [ \pi \, \sqrt{b MR} - b
\, \ln (MR)- b \, (1-\ln 2) \right ]~.
\label{r1}
\ee
Being combined with the expression (\ref{f1}) this yields the formula
\be
{\cal R}=  \sqrt {\eps_1+\eps_2 \over 2 \eps_1} \, {\Gamma(b+1) \over \sqrt{2
\pi \, b}}\,  \exp \left [- {\pi \over 2} \, MR + \pi \, \sqrt{b MR} - b \, \ln
(MR)- b \, (1-\ln 2) \right ]~,
\label{rh}
\ee
which contains an essential dependence on the regulator mass parameter $M$. We
will show however that all such dependence in the phase transition rate can be
absorbed in renormalization of the parameter $\mu$ in the leading semiclassical
term.

\section{Renormalization of $\mu$}
The parameter $\mu$ is defined in the action (\ref{a0}) as the coefficient in
front of  the length of the  boundary between the world sheets for two phases of
the string. Generally this parameter gets renormalized by the quantum
corrections, and in order to find such renormalization at the level of first
quantum corrections, one needs to perform the path integration using the
quadratic part of the action around a configuration, in which the length of the
interface is an arbitrary parameter. For a practical calculation of this effect
we consider a Euclidean space configuration, shown in Fig.3, with the string
lying flat along the $x$ axis, and the interface between the two phases being at
$x=0$. The length of the world line for the boundary is thus the total size $T$
of the world sheet in the time direction. It should be mentioned that such
configuration with different string tension on each side of the boundary is not
stationary for the action (\ref{a0}). However it can be `stabilized' by a source
term (external force) depending on the coordinate $x(t)$ of the boundary: $\int
J(t) x(t) \, dt$, which term does not affect the quadratic in fluctuations part
of the action.

\begin{figure}[ht]
  \begin{center}
    \leavevmode
    \epsfxsize=12cm
    \epsfbox{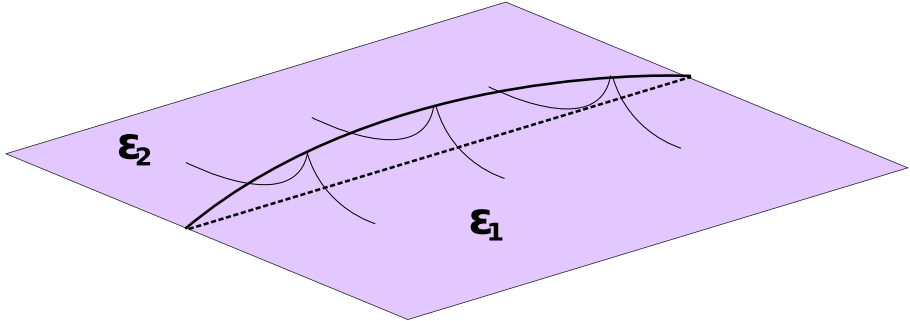}
    \caption{The configuration for the calculation of the renormalization of
$\mu$.}
  \end{center}
\end{figure}

The Gaussian path integral over the transverse coordinates $z(x,t)$ is then to
be calcite with the zero boundary conditions at the edges of the total world
sheet. Using the notation $\zeta(t)$ for the transverse shift of the boundary,
this condition implies $\zeta(0)=\zeta(T)=0$, so that the function $\zeta(t)$
has the Fourier expansion of the form
\be
\zeta(t)=\sqrt{2 \over T} \, \sum_{n=1}^\infty a_n \, \sin \left ( {\pi \, n \,
t \over T} \right )~,
\label{zt}
\ee
and a similar expansion applies to the regulator boundary function $\zeta_R(t)$.
The part of the effective action associated with the boundary is determined by
the functions $z_c(x,t)$ for the transverse shift of the string and the
corresponding regulator functions $Z_c(x,t)$ that satisfy the equations
\be
\Delta z_c=0 ~~~{\rm and} ~~~(\Delta - M^2) \,Z_c =0~,
\label{laphel}
\ee
and the boundary conditions
\be
z_c(0,t)=\zeta(t)~,~~~~Z_c(0,t)=\zeta_R(t)
\label{bcf}
\ee
as well as zero boundary conditions at the edges of the total world sheet. One
can readily find these functions for each harmonic of the boundary values
$\zeta$ and $\zeta_R$, using the normalized to one solutions for $z_c$ and $Z_c$
in each harmonic:
\be
z_c^{(n)}(x,t) = \exp \left ( -|x| \,   {\pi \, n \over T}\right ) \, \sin \left
( {\pi \, n \, t \over T} \right ) ~,
\label{zcf}
\ee
and
\be
Z_c^{(n)}(x,t) = \exp \left [ -|x| \, \sqrt{ \left ( {\pi \, n \over T}\right
)^2 + M^2} \, \right ] \, \sin \left ( {\pi \, n \, t \over T} \right )~.
\label{zrcf}
\ee

In order to separate the boundary effect in the path integral around the
considered configuration from the bulk effects, we again divide it by the path
integral around the configuration where the whole world sheet is occupied by the
same phase of the string. The latter phase can be chosen with either of the
tensions, or with an arbitrary tension $\eps$. Such division results, as
previously, in the cancellation of the bulk contributions, and
the remaining part of the effective action associated with the boundary is
written in
terms of the regularized path integral over the boundary function $\zeta$ as
\bea
&&\mu_R \, T = \mu \, T - \nonumber \\
&&\ln  {\int {\cal D} \zeta  \, \exp \left \{-{1 \over 2} \, \int dt \left [ \mu
\dot{\zeta}^2 + \zeta(t) \, \left ( \eps_2  \, z_c'(x,t)|_{x \to -0} -  \eps_1
\, z_c'(x,t)|_{x \to +0} \right ) \right ] \right \} \over  \int {\cal D}
\zeta_R  \, \exp \left \{ -{1 \over 2} \, \int dt \left [ \mu  \dot{\zeta_R}^2 +
\zeta_R(t) \, \left (\eps_2 \, Z_c'(x,t)|_{x \to -0} -  \eps_1  \, Z_c'(x,t)|_{x
\to +0} \right ) \right ] \right \} } + \nonumber \\
&&\ln  {\int {\cal D} \zeta  \, \exp \left \{- \eps  \, \int dt  \zeta(t) \,
z_c'(x,t)|_{x \to -0}  \right \} \over  \int {\cal D} \zeta_R  \, \exp \left \{
-  \eps \, \int dt   \zeta_R(t) \, Z_c'(x,t)|_{x \to -0}  \right \} }~,
\label{muet}
\eea
where $\mu_R=\mu + \delta \mu$ is the renormalized mass parameter.  The
correction to $\mu$ can thus be written in the form
\be
\delta \mu= -{1 \over 2 \, T} \ln \left \{ \prod_{n=1}^\infty {n^2 +{\tilde b}
\, \sqrt{(MT/\pi)^2+n^2} \over n^2 + {\tilde b} \, n} \, \prod_{n=1}^\infty {n
\over \sqrt{(MT/\pi)^2+n^2}}\right \}~,
\label{ptb}
\ee
where
\be
{\tilde b}= { \eps_1+\eps_2 \over \mu} \, {T \over \pi}~.
\label{tbd}
\ee
In the limit $\eps_1+\eps_2 \ll \mu M$ one can directly apply the results in
Eqs.(\ref{mprod}) and (\ref{r1}) for evaluation of the expression (\ref{ptb})
and write
\bea
\delta \mu &=& -{1 \over 2 \, T} \left ( {\tilde b} \ln {\tilde b} - {\tilde b}
- {M T \over 2} + \pi \sqrt{ {\tilde b} \, {M T \over \pi}} - {\tilde b} \ln {MT
\over \pi} - {\tilde b} \, (1- \ln2) \right )\nonumber \\
&=&{M \over 4} - {1 \over 2} \, \sqrt{(\eps_1+ \eps_2) \, M \over \mu} -
{\eps_1+\eps_2 \over 2 \pi \mu} \, \ln {2 \, (\eps_1+ \eps_2) \over \mu \, M}~,
\label{dmu}
\eea
where a use is made of the Stirling formula
$$\ln {\Gamma({\tilde b}+1) \over \sqrt{2 \pi {\tilde b}}} \to {\tilde b} \,
(\ln {\tilde b} - 1)~,$$
considering that ${\tilde b}$ is proportional to large $T$.

In the limiting case where $\eps_1+\eps_2 \gg \mu M$ the correction $\delta \mu$
vanishes, so that the renormalization effect is negligible.

\section{The result and discussion}
We can now assemble all the relevant elements into a formula for the rate of
the considered transition. Clearly, the path integration over the  variables $z$
factorizes for each of the $(d-2)$ transverse dimensions, so that the expression
for the decay rate takes the form
\be
{d \Gamma \over d \ell}={\eps_1-\eps_2 \over 2 \pi} \,{\cal R}^{d-2} \, \exp
\left ( -
\, {\pi \, \mu^2 \over \eps_1-\eps_2} \right )~,
\label{gfr}
\ee
where $\mu$ is the zeroth order mass parameter. In the case of large string
tension, $\eps_1 + \eps_2 \gg \mu \, M_0$, the `bare' $\mu$ coincides with the
renormalized one, and the factor ${\cal R}$ is equal to one. It can be noted
that
the resulting obvious expression for the rate is also correctly given by
Eq.(\ref{gf}) as soon as the factor $F$ in Eq.(\ref{ff}) is taken in the limit
$\eps_1-\eps_2 \ll \eps_1+\eps_2$: $F \to 1$, which limit, as previously
discussed, is mandated in this case.

In the opposite limit of heavy $\mu$, $\eps_1 + \eps_2 \ll \mu \, M_0$, both the
expression (\ref{rh}) depends on the regulator mass $M$ and the $M$-dependent
renormalization of $\mu$ is essential. Taking into account that each of the
transverse dimensions contributes additively to $\delta \mu$ and expressing in
Eq.(\ref{gfr}) the bare $\mu$ through the renormalized one: $\mu=\m_R-\delta
\mu$, one readily finds that the dependence on the regulator mass M cancels in
the transition rate, and one arrives at the formula given by Eq.(\ref{gf}) and
Eq.(\ref{ff}).

The formula (\ref{ff}) is applicable for arbitrary ratio of the string tensions
$\eps_2/\eps_1$. In particular it can be also applied at $\eps_2 = 0$, in which
case the considered transition describes a complete breaking of the string.

It can be also noted that numerically the factor $F$ depends very moderately on
the ratio of the tensions and changes approximately linearly between $F(0) =
e/\sqrt{4 \pi} = 0.7668\ldots$ and $F(1)=1$. We thus conclude that the
two-dimensional expression (\ref{c2}) for the pre-exponential factor in the
transition rate provides a fairly accurate approximation in higher dimensions as well, as long as the exponential factor is expressed in terms of the physical renormalized mass $\mu$.

There is however an interesting methodical point pertaining to the considered
here problem. Indeed, as was already mentioned, the difference from the problem
of particle creation by external electric field is that the motion of the
ends of the string involves in addition to the mass $\mu$ also an adjacent part
of the string. In terms of the calculation of the path integrals around the
bounce the difference is that the spectrum of soft modes in the particle
creation problem (as well as in that of the two-dimensional false vacuum decay)
consists only of the modes associated with one-dimensional world line of the
boundary of the bounce. The entire pre-exponential factor can then be found
using the effective low energy action for these modes\cite{mv85}. In the
considered here string transition there are also low modes in the bulk of the
world sheet of the string, and there is no parametric separation of their
eigenvalues from those of the modes associated with fluctuations of the
boundary. In the presented calculation the separation of the boundary and bulk
variables is achieved through an `artificial' organization of the normalization
partition function for a flat string into boundary and bulk factors ${\cal
Z}_{\rm boundary}$ and ${\cal Z}_{\rm bulk}$. The bulk contribution then cancels
in the ratio of the partition functions near the bounce and near a flat string,
so that the remaining calculation is reduced to considering the integrals over
the boundary functions only. One can also readily notice that the additional
contribution to the action from the boundary terms as e.g. those with the
functions $z_{1c}$ and $z_{2c}$  in Eq.(\ref{z12by}) corresponds to precisely
the effect of `dragging' of the string by its end.

\section*{Acknowledgment}
The work of M.B.V. is supported in part by the DOE grant DE-FG02-94ER40823.


\begin{thebibliography}{99}

\bibitem{Vilenkin}
  A.~Vilenkin,
  Nucl.\ Phys.\  B {\bf 196}, 240 (1982).
\bibitem{Preskill}
  J.~Preskill and A.~Vilenkin,
  Phys.\ Rev.\  D {\bf 47}, 2324 (1993)
  [arXiv:hep-ph/9209210].
\bibitem{Shifman}
  M.~Shifman and A.~Yung,
  Phys.\ Rev.\  D {\bf 66}, 045012 (2002)
  [arXiv:hep-th/0205025].
\bibitem{vko}
  M.~B.~Voloshin, I.~Y.~Kobzarev and L.~B.~Okun,
  Sov.\ J.\ Nucl.\ Phys.\  {\bf 20}, 644 (1975)
  [Yad.\ Fiz.\  {\bf 20}, 1229 (1974)].
\bibitem{Coleman}
  S.~R.~Coleman,
  Phys.\ Rev.\  D {\bf 15}, 2929 (1977)
  [Erratum-ibid.\  D {\bf 16}, 1248 (1977)].

\bibitem{Schwinger}
  J.~Schwinger, Phys.\ Rev.\ {\bf 86}, 664 (1951).

\bibitem{Callan}
  C.~G.~Callan and S.~R.~Coleman,
  Phys.\ Rev.\  D {\bf 16}, 1762 (1977).
\bibitem{Stone}
  M.~Stone,
  Phys.\ Rev.\  D {\bf 14}, 3568 (1976).
\bibitem{ks}
  V.~G.~Kiselev and K.~G.~Selivanov,
  JETP Lett.\  {\bf 39}, 85 (1984)
  [Pisma Zh.\ Eksp.\ Teor.\ Fiz.\  {\bf 39}, 72 (1984)].
\bibitem{mv85}
  M.~B.~Voloshin,
  Yad.\ Fiz.\  {\bf 42}, 1017 (1985)
  [Sov.\ J.\ Nucl.\ Phys.\  {\bf 42}, 644 (1985)].



\end{thebibliography}
\end{document}